\documentclass[journal]{IEEEtran}

\usepackage{epsfig}
\usepackage{epstopdf}
\usepackage{subfigure}
\usepackage{amsfonts}
\usepackage{amsmath}
\usepackage{amssymb}
\usepackage{graphicx}
\usepackage{epstopdf}
\usepackage{subfigure}
\usepackage{url}
\usepackage{cite}
\usepackage{psfrag}
\usepackage{array}
\usepackage{mdwmath}
\usepackage{mdwtab}
\usepackage{bm}
\usepackage{color}
\usepackage{tabulary}
\usepackage{multirow}
\usepackage{comment}
\usepackage[english]{babel}
\usepackage{algorithm}
\usepackage{algorithmic}
\usepackage{flushend}
\usepackage{caption}
\usepackage{stfloats}

\begin{document}

\title{Streaming Virtual Reality Content}

\author{
\IEEEauthorblockN{Tarek El-Ganainy, Mohamed Hefeeda} \\
\IEEEauthorblockA{School of Computing Science \\ Simon Fraser University \\ Burnaby, BC, Canada}
}

\maketitle

\raggedbottom

\begin{abstract}

The recent rise of interest in Virtual Reality (VR) came with the availability of commodity commercial VR products, such as the Head Mounted Displays (HMD) created by Oculus and other vendors. To accelerate the user adoption of VR headsets, content providers should focus on producing high quality immersive content for these devices. Similarly, multimedia streaming service providers should enable the means to stream 360 VR content on their platforms. In this study, we try to cover different aspects related to VR content representation, streaming, and quality assessment that will help establishing the basic knowledge of how to build a VR streaming system.

\end{abstract}


\section{Introduction}

\IEEEPARstart{I}{ntrest} in Virtual Reality (VR) is on the rise. Till recently, VR was just being studied at universities labs or research institutes with few failed trials to commercialize it. Many obstacles were in the way to provide end users with VR devices. For instance, the size of the headsets used to be huge, the quality of the screens were low, no much content were specifically designed for VR devices, and the accuracy of head tracking wasn't that good which lead to discomfort issues. Over the years, researchers / industry worked on solving those issues. Nothing major was done till the introduction of Oculus Rift \cite{rift}. Thereafter, many of the major players in the computer industry introduced their own headsets. For example: Google Cardboard \cite{GoogleCB} and Daydream \cite{GoogleDD}, HTC VIVE \cite{HTCvive}, Sony PlayStation VR \cite{SonyVR}, Samsung GearVR \cite{SamsungGearVR}. With the increasingly progress of consumer grade VR headsets a tremendous attention is directed to creating and steaming content to such devices. The headsets providers, in addition to many others, are currently producing 360 stereoscopic cameras to enable the creation of VR content. For instance, GoPro Omni \cite{GoProOmni}, Google Odyssey \cite{GoogleOdyssey}, Samsung Project Beyond \cite{SamsungBeyond}, Facebook Surround 360 \cite{FBSurrond}. Similarly, major multimedia streaming service providers such as Facebook \cite{Facebook} and YouTube \cite{YouTube} are currently supporting 360 video streaming for VR devices.

\begin{figure}[H]
    \begin{center}
        \includegraphics[width=.30\textwidth]{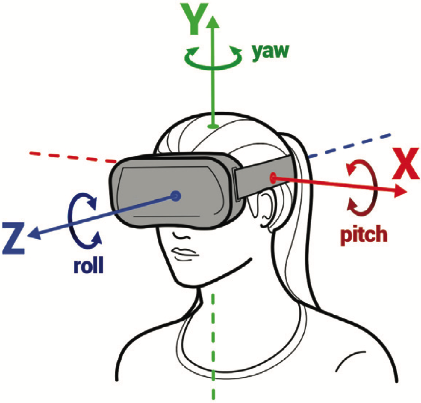}
        \caption{Viewing 360 content on a VR head mounted display. \cite{qian2016optimizing}}     \vspace*{-0.6cm}
        \label{fig:vr_hmd}
   \end{center}
\end{figure}

VR content, a.k.a spherical / 360, can't be viewed with traditional methods. Typically users view VR content on Head Mounted Display (HMD), such as the Oculus Rift. As shown in Fig. (\ref{fig:vr_hmd}), users can move their heads around the immersive 360 space in all possible directions. Head rotations can be simplified using Euler angles (pitch, yaw, roll) which corresponds to rotations around the (x, y, z) axes respectively. The user viewport can be defined as their head rotation angles, and the Field of View (FOV) of the HMD. When the users change their viewport, only the corresponding viewable area of the immersive environment is displayed. That brings us to one of the main challenges in streaming VR content, which is wasting the limited bandwidth while streaming parts of the 360 sphere that won't be viewable by the user. This problem exists in other research domains but is crucial in the context of VR, as the content required to deliver immersive experience is very bandwidth demanding (4K+ resolutions, and up to 60 fps). Intuitively, a solution would be to stream only the user's current viewable area. On the other hand, if the users move their head fast, which is common in VR, they will experience pauses to buffer the new viewport which strongly affects the immersive experience.

Not much work has been done on how to optimize streaming VR content. Nonetheless, there has been work in other fields such as streaming high-resolution-videos for online video lectures, and sports games, where users are allowed to do operations such as zoom / pan. The VR streaming problem can be formulated as the problem of panning around a high-resolution-video using head movements. The major difference between the VR, and the zoom / pan scenarios, is that in the latter the user is more tolerant to buffering time and sudden degradation in quality. In our study, we'll discuss different types of representations for the VR content. Then, we'll try to cover many techniques that have been proposed to improve bandwidth efficiency for high-resolution-videos streaming. Also, we'll go over most of the recent studies dedicated to streaming VR content. Finally, we'll present multiple models to assess the Quality of Experience (QoE) in a VR streaming system.

\section{Content Representation}

VR videos are typically shot using multiple cameras pointing at different directions. The collective field of view for the cameras should cover the 360 space while having enough overlap between them for later stitching purposes. Next, video streams from all the cameras are synchronized, stitched, and projected on a sphere. Finally, to compress the video using standard commercial encoders, we need the video to be in a planar format. Multiple sphere-to-plane mappings have been proposed in that concern \cite{EquirectangularProj}, \cite{FBCubeMap}, \cite{FBOffsetCubeMap}, \cite{FBPyramid}, \cite{li2016novel}, \cite{fu2009rhombic}. Sphere-to-plane mappings can be categorized to two main categories: (1) Uniform Quality Mappings: all parts of the sphere are mapped to the 2D plane with uniform quality, (2) Variable Quality Mappings: more quality is reserved for areas users are currently viewing or more likely to view.

\subsection{Uniform Quality Mappings}

\begin{figure}[H]
    \begin{center}
        \includegraphics[width=.35\textwidth]{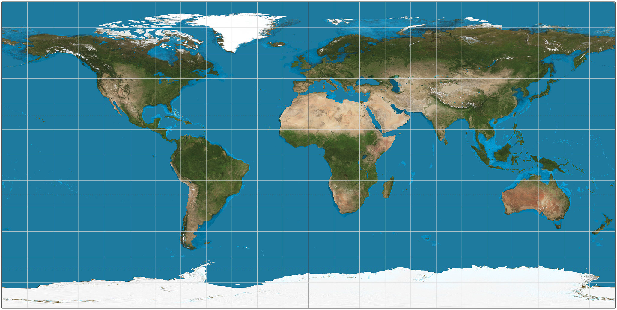}
        \caption{Equirectangular projection of the world map. \cite{EquirectangularProj}.}     \vspace*{-0.6cm}
        \label{fig:equirectangular}
   \end{center}
\end{figure}

\subsubsection{Equirectangular Projection}

is the most common sphere-to-plane mapping. It can be described as unwrapping a sphere on a 2D rectangular plane with the dimensions ($2 \pi r$, $\pi r$), where $r$ is the radius of the sphere. A simple unwrapping of a sphere on a 2D plane will lead to gaps in the output mapping that increase towards the poles. But, for equirectangular projection we stretch the sphere to fit the whole 2D plane. The most known example for equirectangular projection is the world map as shown in Fig. (\ref{fig:equirectangular}). As mentioned above, equirectangular projection is widely supported and easily viewable even with no special players. On the other hand, one of its main drawbacks is the amount of redundant pixels, specially around the poles, which will waste the user's limited bandwidth in a streaming scenario.

\begin{figure}[H]
    \begin{center}
        \includegraphics[width=.48\textwidth]{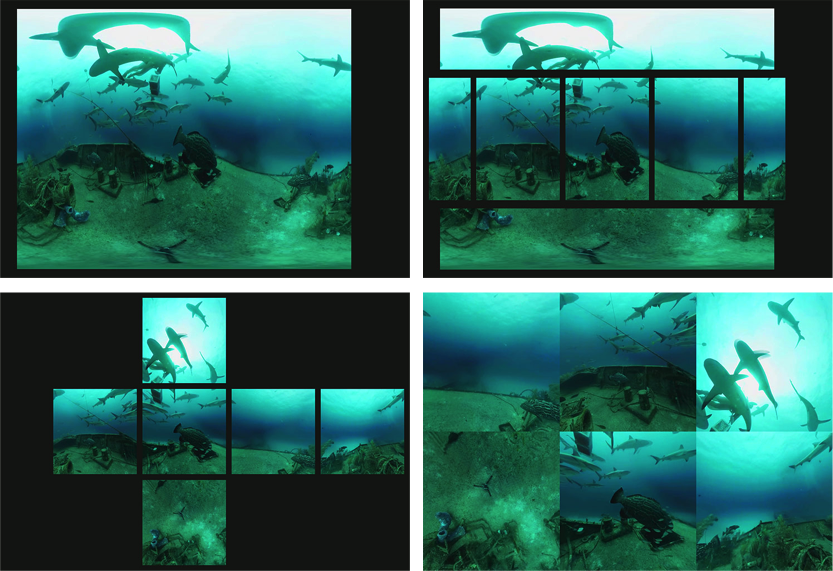}
        \caption{Mapping an equirectangular (spherical) frame to cubemap projection. \cite{FBCubeMap}.}     \vspace*{-0.6cm}
        \label{fig:cubemap}
   \end{center}
\end{figure}

\subsubsection{Cubemap Projection}

has been used extensively in gaming applications, and can be easily rendered using graphics libraries. Similar to equirectangular projection, we are trying to project a sphere on a 2D rectangular plane, but in this case we project the sphere on a cube first. Unlike equirectangular projection, cubemap doesn't impose redundant pixels. A visual example for cubemap projection is shown in Fig. (\ref{fig:cubemap}). Recently, Facebook released an open source implementation for an ffmpeg \cite{FFmpeg} filter to convert a video mapped as an equirectangular to cubemap \cite{FBCubeMapCode}.

\subsubsection{Tile Segmentation Scheme}

\begin{figure}[H]
    \begin{center}
        \includegraphics[width=.48\textwidth]{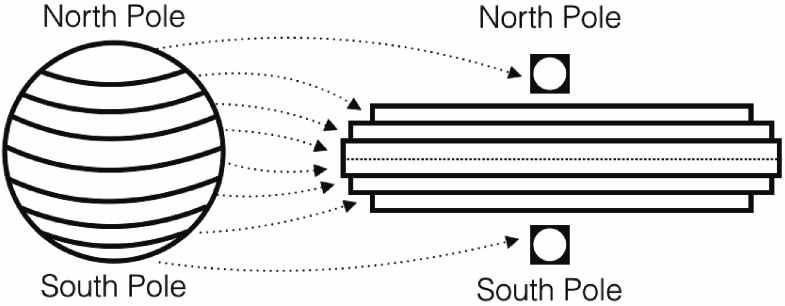}
        \caption{Tile segmentation example with 7 tiles \cite{li2016novel}.}     \vspace*{-0.6cm}
        \label{fig:tiled_segmentation}
   \end{center}
\end{figure}

Li et al. \cite{li2016novel} proposed a sphere-to-plane mapping based on tiling. The main idea is to remove any redundant data as for equirectangular projection. That is, each tile is represented on a plane area that is nearly equal to the area it occupies on the sphere as shown in Fig. (\ref{fig:tiled_segmentation}). They show that their proposed projection can save up to 28\% of the pixel area compared to the traditional equirectangular projection.

\subsubsection{Rhombic Dodecahedron Map}

\begin{figure}[H]
    \begin{center}
        \includegraphics[width=.48\textwidth]{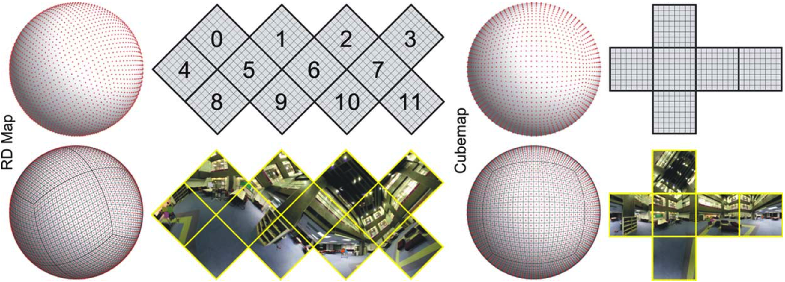}
        \caption{Comparison between rhombic dodecahedron map (left) and cubemap (right) \cite{fu2009rhombic}.}     \vspace*{-0.6cm}
        \label{fig:rd_cubemap}
   \end{center}
\end{figure}

\begin{figure}[H]
    \begin{center}
        \includegraphics[width=.48\textwidth]{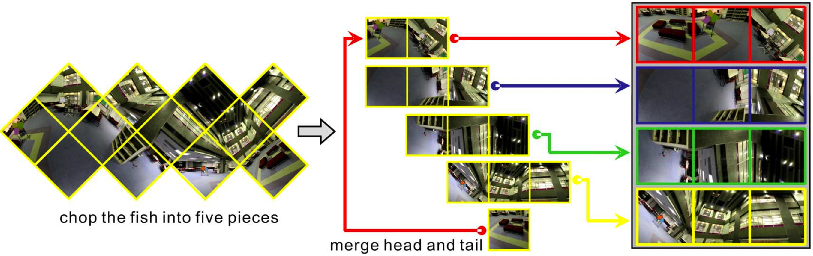}
        \caption{Organizing the fish-like form into MPEG friendly format for encoding. \cite{fu2009rhombic}.}     \vspace*{-0.6cm}
        \label{fig:rd_encoding}
   \end{center}
\end{figure}

Fu et al. \cite{fu2009rhombic} focus on sphere-to-plane mappings, and design a new mapping that offers a better visual quality, stability, compression efficiency. That is, they propose the Rhombic Dodecahedron Map (RD Map) shown in Fig. (\ref{fig:rd_cubemap}), (\ref{fig:rd_encoding}).

\subsection{Variable Quality Mappings}

\begin{figure}[H]
    \begin{center}
        \includegraphics[width=.48\textwidth]{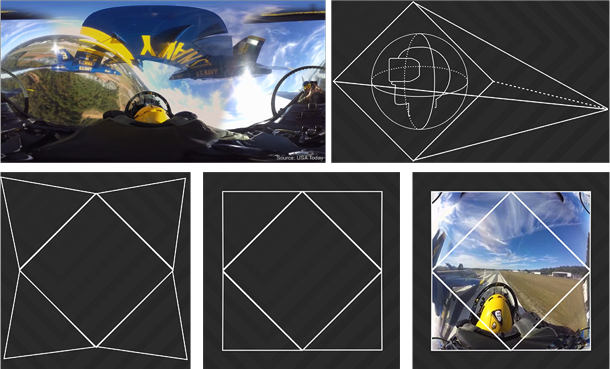}
        \caption{Mapping an equirectangular frame to pyramid projection, then unwrapping the pyramid on a 2D plane.  \cite{FBPyramid}.}     \vspace*{-0.6cm}
        \label{fig:pyramid_proj}
   \end{center}
\end{figure}

\subsubsection{Pyramid Projection}

is one of the early trials by Facebook to support variable quality mappings \cite{FBPyramid}. The main idea is to project the sphere on a pyramid where its base is the user's current viewing area as shown in Fig. (\ref{fig:pyramid_proj}). By doing so, the user's viewport will be represented with highest number of pixels, and we'll have a degradation of quality as the user moves his head to the back. There are two main issues with pyramid projection: (1) as the users rotate their head by $120^{\circ}$, the quality drops the same amount as they turn their head to the back of the pyramid, (2) Since pyramid projection is not supported on GPUs, it's not as efficient to render them as it is for a cubemap.

\begin{figure}[H]
    \begin{center}
        \includegraphics[width=.35\textwidth]{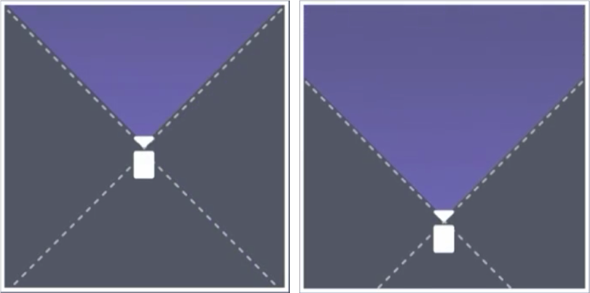}
        \caption{cubemap (left) and offset-cubemap (right) \cite{FBOffsetCubeMap}.}     \vspace*{-0.6cm}
        \label{fig:offset_cubemap}
   \end{center}
\end{figure}

\subsubsection{Offset-cubemap Projection}

builds on top of the cubemap to provide a variable quality mapping while addressing the issues of the pyramid projection. Offset-cubemap is a regular cubemap where the user is pushed back from the center of the cube with a predefined offset in the opposite direction from where he is looking as shown in Fig. (\ref{fig:offset_cubemap}). Since offset-cubemap is essentially a cubemap, we can efficiently render it on a GPU. Also, offset-cubemaps offer smoother degradation of quality as the user moves his head around to the back compared to a pyramid projection. Offset-cubemap projection has been proposed recently in Facebook F8 developers conference \cite{FBOffsetCubeMap}, but no proper documentation has been released for it yet. As a result, it hasn't been widely adopted, even a recent measurement study \cite{qian2016optimizing} on Facebook's 360 streaming pipeline shows that they are still using the traditional cubemap projection.

\section{Tiled / ROI Streaming}

In the last few years, we noticed an increase in the availability of high-resolution-videos. Examples for high-resolution-videos are: panoramic and spherical videos, which are the focus of our study. Streaming such content to users over the Internet demands very high bandwidth connections. A typical scenario for streaming high-resolution-videos is online video lectures, or streaming 360 VR videos. In such scenarios, the user usually views only a portion of the video at a time. As a result, there is a huge waste of bandwidth for streaming content that is not visible to the user. This issue has been addressed in many studies \cite{quang2010supporting}, \cite{wang2014mixing}, \cite{de2016efficient}. That is, researchers try to optimize streaming high-resolution-videos to users while maintaining a good quality of experience. Some of the works utilize the concept of Region of Interest (ROI) \cite{quang2010supporting}, \cite{ngo2011adaptive}. By knowing the user ROIs we can stream it with high quality while minimizing the quality of the rest of the video and saving the user bandwidth. Others tackle the problem of streaming high-resolution-videos over limited bandwidth by tiling \cite{d2016using}, \cite{zare2016hevc}. That is, the video is partitioned to multiple tiles and depending on the user's viewable area, we stream the overlapping tiles. Tiling has proven effective in domains such as online video lectures \cite{makar2010real}, and sports \cite{gaddam2016tiling}. Furthermore, other issues imposed by tiled / ROI streaming have been discussed, as encoding performance \cite{makar2010real}, blending tiles with different qualities \cite{yu2015content}, and stitching tiles on the client side \cite{sanchez2015compressed}.

\subsection{Challanges}

One of the most important challenges of ROI streaming is how to crop them dynamically in a fast and scalable way. We can think of some naive solutions like cropping all the possible ROIs from the video and encoding them independently. One major drawback of such approach is that it will consume a big amount of storage. A solution could be to only process the popular ROIs based on user access patterns. Such solution will reduce the flexibility of the system when users request ROIs that weren't processed before. Another naive flexible solution is to process ROIs on the fly. Once the server receives the coordinates of the ROI from the client, it crops the appropriate segment from the video, encodes it, and transmits it back to the client. Although this solution is flexible enough to support any ROI, it cannot scale well with the increase in number of users.

\begin{figure}[H]
    \begin{center}
        \includegraphics[width=.48\textwidth]{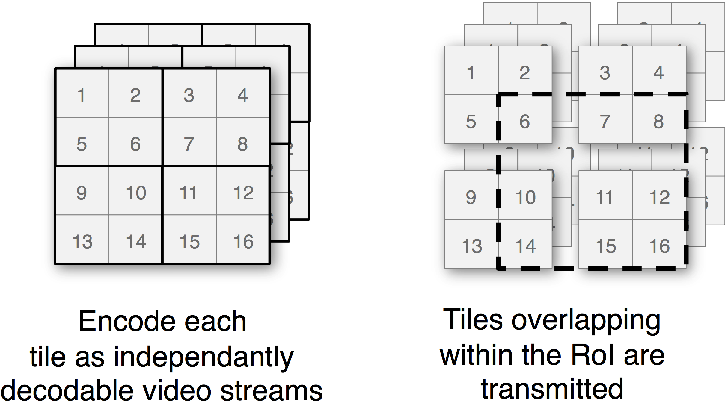}
        \caption{Tiled Streaming. Each tile is encoded independently. Tiles overlapping with requested ROI sent to client for decoding. \cite{ngo2011adaptive}}     \vspace*{-0.6cm}
        \label{fig:tiled_streaming}
   \end{center}
\end{figure}

Khiem et al. \cite{quang2010supporting} proposed two schemes supporting ROI streaming, while encoding the video only once. In the first scheme, namely tiled streaming, each video is divided into a grid of tiles encoded independently. When the client requests a ROI, the server responds with the tiles overlapping with the ROI requested as shown in Fig. (\ref{fig:tiled_streaming}).

\begin{figure}[H]
    \begin{center}
        \includegraphics[width=.48\textwidth]{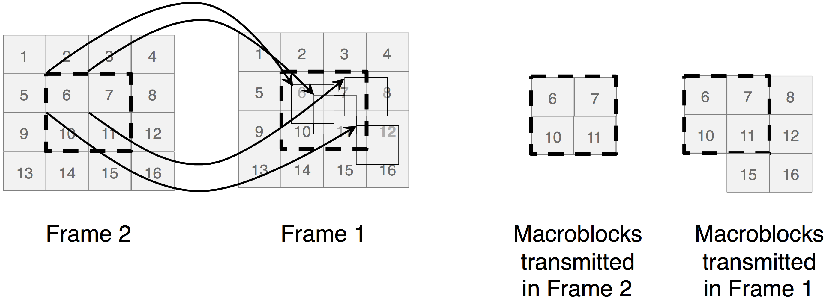}
        \caption{Monolithic Streaming. All macro-blocks belonging to the ROI along with dependent macro-blocks needed for its decoding are sent to the client.  \cite{ngo2011adaptive}}     \vspace*{-0.6cm}
        \label{fig:monolithic_streaming}
   \end{center}
\end{figure}

In the second scheme, namely monolithic streaming, each video is encoded only once as one entity. When a client requests a ROI, the server responds with the macro-blocks containing the ROI and all other dependent macro-blocks needed to decode them as shown in Fig. (\ref{fig:monolithic_streaming}).

Khiem et al. tackled the scalability and flexibility issues discussed earlier for both the tiled and monolithic streaming. On the other hand, their approach introduced a new problem of sending extra data that are needed for transmission but not viewable to the users. That is, for tiled streaming, some of the tiles sent don't lie in the ROI requested. Similarly, for monolithic streaming, dependent macro-blocks aren't viewable by the user. To address the extra data sent problem, they studied the trade-off between the compression ratio and minimizing the tile size for tiled streaming, or reducing the motion vectors search space for monolithic streaming. That is, when we minimize the tile size or reduce the motion vectors search space the compression ratio drops, but less extra data is sent to the client.

In their later work, Khiem et al. \cite{ngo2011adaptive} try to further reduce the extra data from being streamed to the clients. They realize that users tend to have patterns viewing ROIs, and that not all of them are equally important. Their method is based on encoding different parts of the video with different encoding parameters based on their popularity. For instance, in tiled streaming, tiles lying within a popular ROI will be smaller in size compared to less popular areas. Also, for monolithic streaming, the motion vector search space for macro-blocks lying within a popular ROI will be more constrained compared to a wider search range for less popular ROIs. They show that by applying these optimizations they can reach up to 21\%, 27\% bandwidth reduction using monolithic and tiled streaming respectively.

As discussed above, tiled / ROI methods have been shown effective in the context of streaming high-resolution-videos, yet they have some practical issues that researchers are trying to address:
\begin{itemize}
\item Encoding performance: tiled / ROI methods usually require multi-pass encoding for each video, which is computationally costly. This issue turns to be critical for large scale system.
\item Stitching / blending: for tiled streaming, the client will need to have an additional component to join the tiles before decoding, or use multiple decoders.
\item Frequency of adaptation: user interest may evolve over time, meaning that the user may focus on different parts of the video when viewing it for another time
\item Different user profiles: different clusters of users may focus on different regions of the videos. For instance, in a sports video,  the interesting regions in the video would vary depending on the team a user supports.
\end{itemize}

\subsection{Encoding Performance}

\begin{figure}[H]
    \begin{center}
        \includegraphics[width=.35\textwidth]{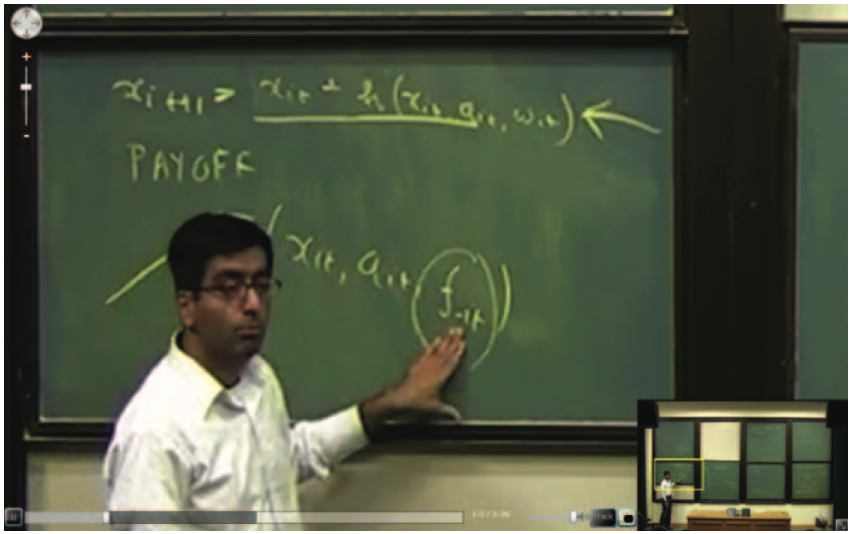}
        \caption{Adaptive ROI streaming client with thumbnail viewer in the bottom right corner. \cite{makar2010real}}     \vspace*{-0.6cm}
        \label{fig:thumbnail_view}
   \end{center}
\end{figure}

Makar et al. \cite{makar2010real} worked on improving the encoding performance of tiled high-resolution-videos in a real time streaming scenario. They make use of the mode and motion vector information available in the original encoded video. Each video is encoded with multiple resolutions, the lowest one is displayed to the user as a thumbnail. For the high resolution videos, they are cropped to multiple tiles and encoded as independent videos. The user can choose the ROI he is interested in viewing from the thumbnail video, as shown in Fig. (\ref{fig:thumbnail_view}). When the server receives a request with a ROI coordinates, it responds with the corresponding overlapping high resolution tiles. Later, if the user changes the viewable ROI, the client requires new tiles to be streamed. Meanwhile, the client up-samples the thumbnail to conceal the missing area till it streams the needed tiles. As a result, they offer the user a seamless experience viewing the video, with minor fluctuations in the perceived quality.

Their approach relies on three main observations. First, the ROI tiles requested by all users do overlap in most cases. So, they can just encode popular tiles offline, while encoding other tiles on demand. Second, based on the nature of the online video lectures scenarios they choose, a static background usually occupies big portion of the video. Given that, while downsampling the original video to lower resolution versions, they check if a macro-block is a static background they just copy it from the previous frame instead of spending time to down-sample it. Lastly, they utilize the motion estimation already performed while encoding the original video. They show that their proposed method can achieve a significant reduction in the encoding time at the server up to 90\% in some cases while not affecting the quality. One of the main drawbacks of their work is that they assume a static camera where the scene comprises low motion. This assumption would work perfectly for scenarios such as surveillance systems or online video lectures, but won't hold strong for immersive VR applications. In case of VR applications, we expect to have highly dynamic scenes. Also, their method requires having multiple decoders running on the client side.

\subsection{MPEG-DASH Spatial Relationship Description}

In a more recent work, D'Acunto et al. \cite{d2016using} make use of the MPEG-DASH Spatial Relationship Description (SRD) \cite{niamut2016mpeg} extensions to support tiled streaming. SRD can describe a video as a spatial collection of synchronized videos. Using SRD they can stream to the users the area they are currently viewing in the highest quality possible based on their available bandwidth. Also, they provide the users with a seamless experience even if they zoom in or pan around. For zoom, the current streamed segment is up-sampled, and displayed to the user while downloading the high quality segment. For pan, they always stream a fallback lowest quality full sized version of the segment, so they can display the corresponding panned area in low quality while downloading the high quality one.

Similarly, Le Feuvre et al. \cite{le2016tiled} utilized the MPEG-DASH SRD extensions to perform tiled streaming. They developed an open-source MPEG-DASH tile-based streaming client using the SRD extensions. They experimented with different adaption modes, both independent H.264 tiles and constrained HEVC tiles.

\subsection{Mixing Tile Resolutions}

Since our focus in this study is immersive VR multimedia experiences, streaming only a portion of the video using tiled / ROI methods won't be sufficient. That is, to provide an immersive experience we need to provide the users with the whole 360 environment around them not to break the sense of presence. To do that while keeping the bandwidth requirement in consideration, we may mix tiles with different resolutions / qualities based on their importance from the user viewing perspective.

Wang et al. \cite{wang2014mixing} studied the effect of mixing tile resolutions on the quality perceived by the users. They conducted a psychophysical study with 50 participants. That is, they show the users tiled videos, where the tile resolutions are chosen randomly from two different levels. The two levels of resolutions are chosen every time so one of them is the original video resolution, and the other is one level lower. The experiments show that in most cases when they mix HD (1920x1080p) tiles with (1600x900p) tiles participants won't notice any difference. Also, when they mix the HD tiles with (960x540p) the majority of the participants are fine with the degradation in quality when viewing low to medium motion videos. Furthermore, for high motion video, more than 40\% of the participants accept the quality degradation.


Zare et al. \cite{zare2016hevc} proposed an HEVC-compliant tiles streaming approach utilizing the motion-constrained tile sets (MCTS) concept. MCTS have been used where multiple tiles are merged at the client side to support devices with a single hardware decoder. They propose to have two versions of the video, one in high resolution, and the other in low resolution. The two versions of the video are partitioned to tiles, and streamed to users based on their current viewport. That is, the tiles currently viewed by the user are streamed in high resolution, while the rest of the tiles are streamed in low resolution. While encoding each independent tile, they allow motion vectors to point to other tiles that are usually streamed together. Multiple tiling schemes are examined, while smaller tiles minimize the extra non-viewable data sent, they offer less compression ratio. The results show that they can achieve from 30\% to 40\% bitrate reduction based on the tiling scheme chosen.


As mentioned before, one of the main challenges of tiled streaming is having multiple decoders at the client side to decode each independent tile. Sanchez et al. \cite{sanchez2015compressed} addressed this challenge to support devices having a single hardware decoder. Their work is based on the low complexity compressed domain stitching process proposed in \cite{ngo2011adaptive}. Each independently encoded tile received by the client is converted to an HEVC tile in the output bitstream. This can be done by fulfilling a set of constraints while encoding the original video. They also propose a method to reduce the transmission bitrate when users switch between ROIs. They show that the bitrate savings vary depending on the video content and ROI switching speed.

\begin{figure}
\centering
\subfigure[]{\label{fig:before_blending}\includegraphics[width=.35\textwidth]{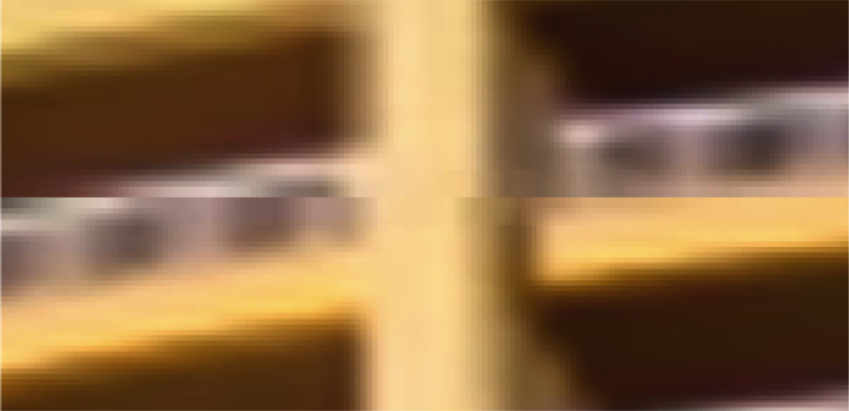}}
\subfigure[]{\label{fig:after_blending}\includegraphics[width=.35\textwidth]{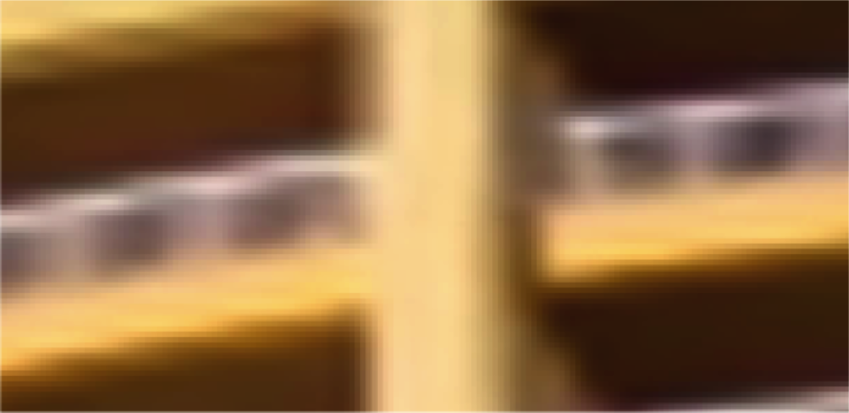}}
\caption{(a) magnified view of the seam resulting from the boundary between two tiles having different resolutions. (b) no visible seam after applying alpha blending on overlapping tiles. \cite{yu2015content} } \label{fig:blending}
\end{figure}

For tiled streaming, typically we partition the video into tiles in both the horizontal and vertical directions while assigning uniform resolution levels to them. Later, we stream only the tiles overlapping with the user viewport. In some studies, mixing tiles with different levels of resolutions is considered to provide a full delivery for spherical / panoramic content. The methods employed to choose the tiles resolution levels were rather crude. Moreover, rendering neighboring tiles with different resolutions will lead to obvious seams that will affect the perceived quality by the users. Yu et al. \cite{yu2015content} focus on the issues mentioned above in their work. First, they partition the equirectangular video to horizontal tiles. Each horizontal tiles is assigned a sampling weight based on it's content and users viewing history. Then, based on the bandwidth budget and the sampling weight of each tile, they optimize the bit-allocation for each tile. Secondly, to overcome the seams problem, they add an overlapping margin between each two neighboring tiles. Finally, alpha blending is applied on the overlapping tile margins to reduce the visible seams effect as shown in Fig. (\ref{fig:blending}).

\subsection{Personalized Views}

Most of the tiled streaming methods entail problems like the need for multiple decoders at the client or sending extra bits from tiles that won't be viewable by the user. De Praeter \cite{de2016efficient} tackle these problems by sending each user a personalized view of the video. Encoding a personalized view for each user is computationally costly and doesn't scale well with the increase in number of users. To address this issue, they accelerate the encoding of each personalized view using information extracted from a pre-analysis done on the original high-resolution-video using HEVC Test Model (HM). They show that by applying their approach they can reduce the complexity of generating each personalized view by more than 96.5\% compared to independently encoding them. This performance boost comes at the cost of bitrate overhead up to 19.5\%. They show that the bitrate overhead resulting from their method is still smaller compared to the one resulting from tiled-based methods.

\section{Streaming Systems}

\subsection{Partial Delivery Systems}

Most of the tile-based streaming systems encode tiles independently at uniform bitrates. This may lead to waste of bandwidth, specially for high bitrate static tiles that could have been streamed with less bitrate and similar quality. Inoue et al. \cite{inoue2010interactive} tackle the aforementioned issue in their system. They propose a tile-based adaptive rate adaptation system using H.264 multiple view MVC standard. Each tile in the video is encoded at multiple bitrates and synchronously multiplexed. That is, for each tile they calculate a quality value and associate it to a $view\_id$ with the corresponding bitrate, and field of view. Subjective experiments show their system can provide higher quality compared to uniform bitrate tile-based systems. Specifically, their system ensures tiles having dense motion to be streamed with the highest possible quality, which in-turn affects user perception dramatically.

\subsection{Full Delivery Systems}

\begin{figure}[H]
    \begin{center}
        \includegraphics[width=.48\textwidth]{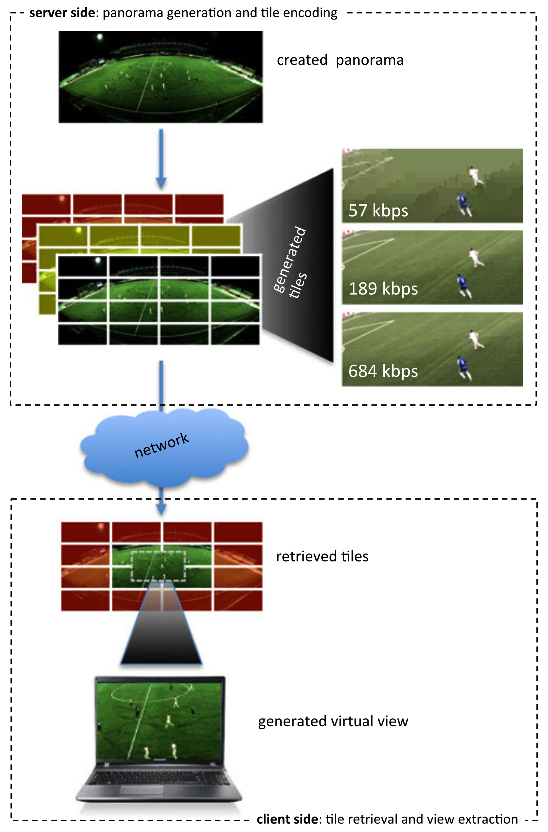}
        \caption{Overview of Gaddam et al. \cite{gaddam2016tiling} panorama streaming system.}     \vspace*{-0.6cm}
        \label{fig:gaddam_system}
   \end{center}
\end{figure}

Gaddam et al. \cite{gaddam2016tiling} developed a streaming system for panoramic videos based on tiling methods. Overview of their system is shown in Fig. (\ref{fig:gaddam_system}).

\begin{figure*}[tp]
\centering
\subfigure[Binary.]{\label{fig:tiled_binary}\includegraphics[width=.35\textwidth]{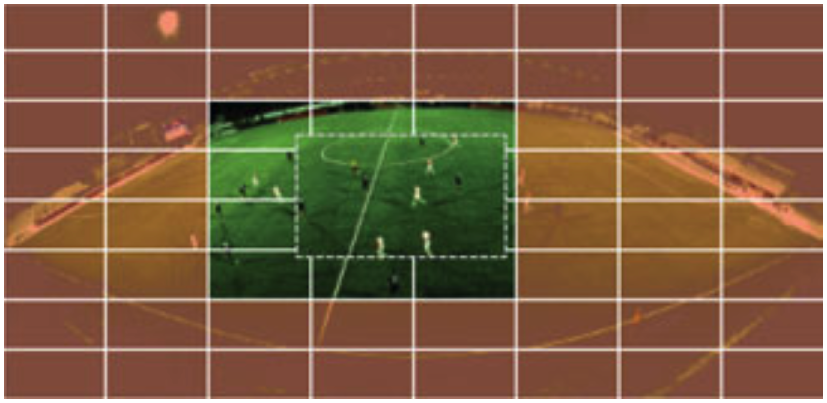}}
\subfigure[Thumbnail.]{\label{fig:tiled_thumbnail}\includegraphics[width=.35\textwidth]{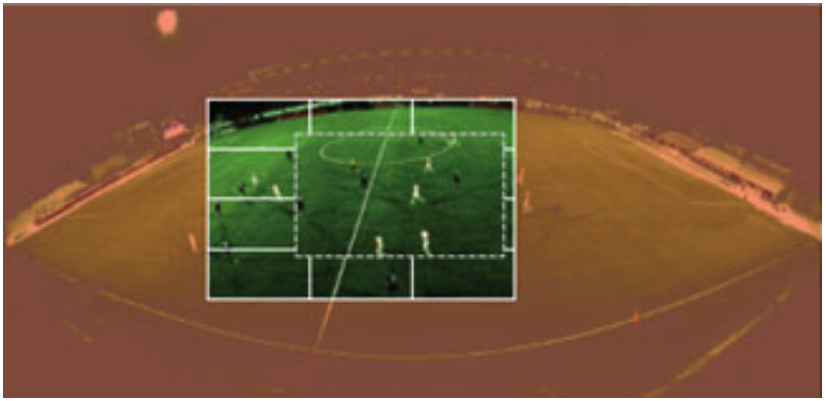}}
\subfigure[Predicted.]{\label{fig:tiled_prediction}\includegraphics[width=.35\textwidth]{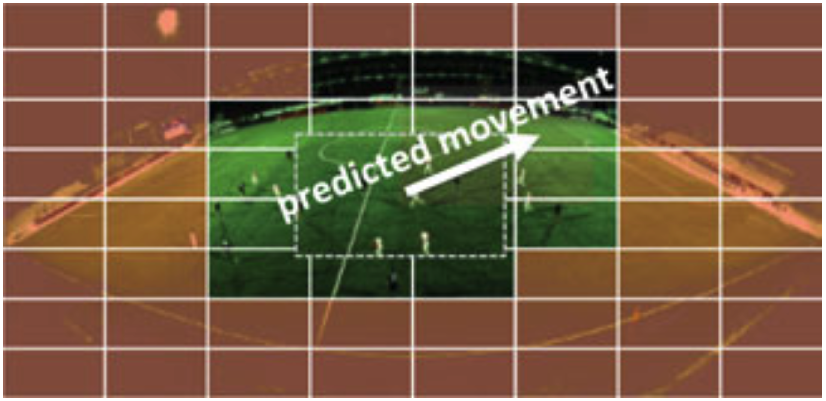}}
\subfigure[Pyramid.]{\label{fig:tiled_pyramid}\includegraphics[width=.35\textwidth]{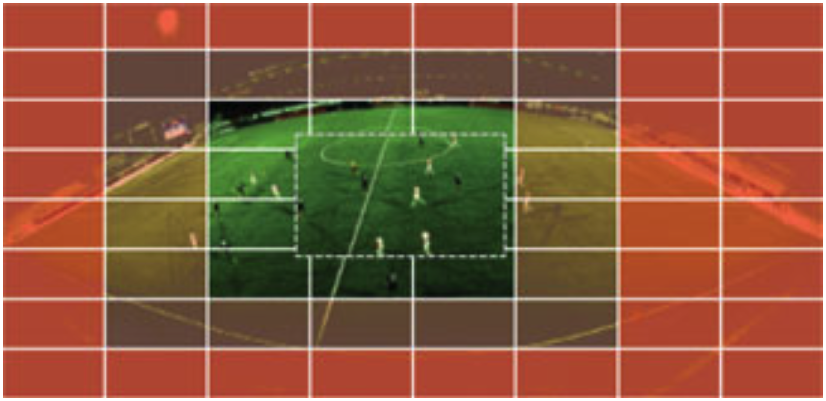}}
\caption{Different tiling schemes proposed by Gaddam et al. \cite{gaddam2016tiling} } \label{fig:tiled_gaddam}
\end{figure*}

They exploit different tiling schemes in their system:

\begin{itemize}
\item Binary: stream only tiles overlapping with the user viewport in high quality, as shown in Fig. (\ref{fig:tiled_gaddam}-a).
\item Thumbnail: stream tiles overlapping with the user viewport in high quality, while streaming the whole video in low quality as shown in Fig. (\ref{fig:tiled_gaddam}-b). If the user switches his viewport, they upscale the thumbnail to avoid buffering time while downloading the high quality tiles of the new viewport.
\item Predicted: stream the user viewport tiles in high quality, similarly, tiles that the user is expected to view next. An example for predicted tiling is shown in Fig. (\ref{fig:tiled_gaddam}-c).
\item Pyramid: stream the user's viewport tiles with the highest quality, while streaming the rest of the tiles with a gradual degradation in quality as shown in Fig. (\ref{fig:tiled_gaddam}-d). That is, the further we move from the center of the user's viewport we stream tiles with less resolution.
\end{itemize}

They evaluate their system using subjective and objective measures. The results show their system can provide similar QoE using pyramidal tiled streaming methods compared to streaming the full quality video while reducing the bitrate.

\subsection{Predictive Systems}

Alface et. al \cite{rondao2012interactive} studied how to improve the quality perceived by optimizing the quality of the user's viewport taking in consideration the bandwidth constraint. Their work depends on estimating the likelihood of users requesting certain viewports. That is, the quality of each region is modulated across different regions in the whole video based on the likelihood that it will be viewed by the user. They show that their system outperforms direct ROI tiling approaches.

In a more recent work, Qian et al. \cite{qian2016optimizing} stream only the visible portion of the video based on head movement prediction. They collected head tracking traces for 5 users. Prediction is done using averaging, linear regression, and average linear regression models. They managed to predict the user head movements with 90\% accuracy using linear regression. Their results show that using traces from users head movements, they can reduce the bandwidth consumption up to 80\%. Moreover, they do a measurement study on commercial 360 video streaming platforms, like Youtube and Facebook. They show that both platforms use standard H.264 encoding in an mp4 container. For the sphere-to-plane mappings, Youtube uses equirectangular projection, while Facebook implements cubemap projection.

\section{Quality Assessment}

In our study, we are focused on spherical (360) videos. Typically, a 360 video is mapped onto one or more planes to be compatible with the current commercial encoders. There haven't been much attention in the literature on how to evaluate the quality of such content. It is not clear how to compare 360 videos under different projections at different bitrates with the original high quality video. For instance, in a typical streaming scenario, when a user changes the viewport, we may render a low quality version of the new viewport while downloading a better quality version, that needs to be accounted for. Also, user access patterns should be incorporated to evaluate the overall quality for different viewports generated for a 360 video.

Yu et al. \cite{yu2015framework} investigate how to assess the quality of 360 videos under different projections and evaluate their coding efficiency. Using head tracking traces, they generate different viewports and use them to compare the original and the constructed 360 videos. Then, they develop a sphere based peak signal-to-noise-ratio (S-PSNR) to approximate the average quality for all possible viewports. Since users tend to view similar parts of the videos, they use head tracking traces to compute a weighted S-PSNR. Finally, they generalize the weighted S-PSNR by assuming it as an approximate for the average viewport PSNR in case we don't have head tracking traces.

Zakharchenko et al. \cite{zakharchenko2016quality} propose an objective quality estimation method for spherical videos. They use a weighted PSNR for equirectangular content and special zero area distortion projection method to compare the constructed and the original views. The reliability and accuracy of their method have been verified by showing good correlation with subjective quality assessments done by a group of experts.

\section{Conclusion}

In this paper, we discussed different parts of a VR streaming system. First, we explained different ways of how to represent spherical content in a compatible way with standard encoders. Then, we discussed different solutions for streaming high-resolution-videos under limited bandwidth conditions. Many of the related works discussed can open up ideas on how to optimize the current VR streaming systems. Next, we showed recent attempts for VR streaming systems. Finally, we presented multiple models that can be used to asses the QoE for a VR streaming system. We think that this survey will help people interested in building VR streaming systems with the basic understanding of its components.

{\tiny
	\bibliographystyle{IEEEtran}
	\bibliography{./ref}

\begin{thebibliography}{10}
\providecommand{\url}[1]{#1}
\csname url@samestyle\endcsname
\providecommand{\newblock}{\relax}
\providecommand{\bibinfo}[2]{#2}
\providecommand{\BIBentrySTDinterwordspacing}{\spaceskip=0pt\relax}
\providecommand{\BIBentryALTinterwordstretchfactor}{4}
\providecommand{\BIBentryALTinterwordspacing}{\spaceskip=\fontdimen2\font plus
\BIBentryALTinterwordstretchfactor\fontdimen3\font minus
  \fontdimen4\font\relax}
\providecommand{\BIBforeignlanguage}[2]{{%
\expandafter\ifx\csname l@#1\endcsname\relax
\typeout{** WARNING: IEEEtran.bst: No hyphenation pattern has been}%
\typeout{** loaded for the language `#1'. Using the pattern for}%
\typeout{** the default language instead.}%
\else
\language=\csname l@#1\endcsname
\fi
#2}}
\providecommand{\BIBdecl}{\relax}
\BIBdecl

\bibitem{rift}
``Oculus rift,'' \url{https://www3.oculus.com/en-us/rift/}, accessed:
  2016-12-06.

\bibitem{GoogleCB}
``Google cardboard,'' \url{https://vr.google.com/cardboard/}, accessed:
  2016-12-06.

\bibitem{GoogleDD}
``Google daydream,'' \url{https://vr.google.com/daydream/}, accessed:
  2016-12-06.

\bibitem{HTCvive}
``Htc vive,'' \url{https://www.vive.com/ca/}, accessed: 2016-12-06.

\bibitem{SonyVR}
``Sony playstation vr,''
  \url{https://www.playstation.com/en-ca/explore/playstation-vr/}, accessed:
  2016-12-06.

\bibitem{SamsungGearVR}
``Samsung gearvr,'' \url{http://www.samsung.com/global/galaxy/gear-vr/},
  accessed: 2016-12-06.

\bibitem{GoProOmni}
``Gopro omni,'' \url{https://vr.gopro.com}, accessed: 2016-12-06.

\bibitem{GoogleOdyssey}
``Google odyssey,'' \url{https://gopro.com/odyssey}, accessed: 2016-12-06.

\bibitem{SamsungBeyond}
``Samsung project beyond,'' \url{http://thinktankteam.info/beyond/}, accessed:
  2016-12-06.

\bibitem{FBSurrond}
``Facebook surround 360,''
  \url{https://facebook360.fb.com/facebook-surround-360/}, accessed:
  2016-12-06.

\bibitem{Facebook}
``Facebook,'' \url{https://www.facebook.com}, accessed: 2016-12-06.

\bibitem{YouTube}
``Youtube,'' \url{https://www.youtube.com}, accessed: 2016-12-06.

\bibitem{qian2016optimizing}
F.~Qian, L.~Ji, B.~Han, and V.~Gopalakrishnan, ``Optimizing 360 video delivery
  over cellular networks,'' in \emph{Proceedings of the 5th Workshop on All
  Things Cellular: Operations, Applications and Challenges}.\hskip 1em plus
  0.5em minus 0.4em\relax ACM, 2016, pp. 1--6.

\bibitem{EquirectangularProj}
``Equirectangular projection,''
  \url{https://en.wikipedia.org/wiki/Equirectangular_projection}, accessed:
  2016-12-06.

\bibitem{FBCubeMap}
``Under the hood: building 360 video,''
  \url{https://code.facebook.com/posts/1638767863078802/under-the-hood-building-360-video/},
  accessed: 2016-12-06.

\bibitem{FBOffsetCubeMap}
``Optimizing 360 video for oculus,''
  \url{https://developers.facebook.com/videos/f8-2016/optimizing-360-video-for-oculus/},
  accessed: 2016-12-06.

\bibitem{FBPyramid}
``Next-generation video encoding techniques for 360 video and vr,''
  \url{https://code.facebook.com/posts/1126354007399553/next-generation-video-encodin…},
  accessed: 2016-12-06.

\bibitem{li2016novel}
J.~Li, Z.~Wen, S.~Li, Y.~Zhao, B.~Guo, and J.~Wen, ``Novel tile segmentation
  scheme for omnidirectional video,'' in \emph{Image Processing (ICIP), 2016
  IEEE International Conference on}.\hskip 1em plus 0.5em minus 0.4em\relax
  IEEE, 2016, pp. 370--374.

\bibitem{fu2009rhombic}
C.-W. Fu, L.~Wan, T.-T. Wong, and C.-S. Leung, ``The rhombic dodecahedron map:
  An efficient scheme for encoding panoramic video,'' \emph{IEEE Transactions
  on Multimedia}, vol.~11, no.~4, pp. 634--644, 2009.

\bibitem{FFmpeg}
``Ffmeg,'' \url{https://www.ffmpeg.org}, accessed: 2016-12-06.

\bibitem{FBCubeMapCode}
``Facebook cubemap transform open source code,''
  \url{https://github.com/facebook/transform}, accessed: 2016-12-06.

\bibitem{quang2010supporting}
K.~Q.~M. Ngo, R.~Guntur, A.~Carlier, and W.~T. Ooi, ``Supporting zoomable video
  streams with dynamic region-of-interest cropping,'' in \emph{Proceedings of
  the first annual ACM conference on Multimedia systems}.\hskip 1em plus 0.5em
  minus 0.4em\relax ACM, 2010, pp. 259--270.

\bibitem{wang2014mixing}
H.~Wang, V.-T. Nguyen, W.~T. Ooi, and M.~C. Chan, ``Mixing tile resolutions in
  tiled video: A perceptual quality assessment,'' in \emph{Proceedings of
  Network and Operating System Support on Digital Audio and Video
  Workshop}.\hskip 1em plus 0.5em minus 0.4em\relax ACM, 2014, p.~25.

\bibitem{de2016efficient}
J.~De~Praeter, P.~Duchi, G.~Van~Wallendael, J.-F. Macq, and P.~Lambert,
  ``Efficient encoding of interactive personalized views extracted from
  immersive video content,'' in \emph{Proceedings of the 1st International
  Workshop on Multimedia Alternate Realities}.\hskip 1em plus 0.5em minus
  0.4em\relax ACM, 2016, pp. 25--30.

\bibitem{ngo2011adaptive}
K.~Q.~M. Ngo, R.~Guntur, and W.~T. Ooi, ``Adaptive encoding of zoomable video
  streams based on user access pattern,'' in \emph{Proceedings of the second
  annual ACM conference on Multimedia systems}.\hskip 1em plus 0.5em minus
  0.4em\relax ACM, 2011, pp. 211--222.

\bibitem{d2016using}
L.~D'Acunto, J.~van~den Berg, E.~Thomas, and O.~Niamut, ``Using mpeg dash srd
  for zoomable and navigable video,'' in \emph{Proceedings of the 7th
  International Conference on Multimedia Systems}.\hskip 1em plus 0.5em minus
  0.4em\relax ACM, 2016, p.~34.

\bibitem{zare2016hevc}
A.~Zare, A.~Aminlou, M.~M. Hannuksela, and M.~Gabbouj, ``Hevc-compliant
  tile-based streaming of panoramic video for virtual reality applications,''
  in \emph{Proceedings of the 2016 ACM on Multimedia Conference}.\hskip 1em
  plus 0.5em minus 0.4em\relax ACM, 2016, pp. 601--605.

\bibitem{makar2010real}
M.~Makar, A.~Mavlankar, P.~Agrawal, and B.~Girod, ``Real-time video streaming
  with interactive region-of-interest,'' in \emph{2010 IEEE International
  Conference on Image Processing}.\hskip 1em plus 0.5em minus 0.4em\relax IEEE,
  2010, pp. 4437--4440.

\bibitem{gaddam2016tiling}
V.~R. Gaddam, M.~Riegler, R.~Eg, C.~Griwodz, and P.~Halvorsen, ``Tiling in
  interactive panoramic video: Approaches and evaluation,'' \emph{IEEE
  Transactions on Multimedia}, vol.~18, no.~9, pp. 1819--1831, 2016.

\bibitem{yu2015content}
M.~Yu, H.~Lakshman, and B.~Girod, ``Content adaptive representations of
  omnidirectional videos for cinematic virtual reality,'' in \emph{Proceedings
  of the 3rd International Workshop on Immersive Media Experiences}.\hskip 1em
  plus 0.5em minus 0.4em\relax ACM, 2015, pp. 1--6.

\bibitem{sanchez2015compressed}
Y.~Sanchez, R.~Skupin, and T.~Schierl, ``Compressed domain video processing for
  tile based panoramic streaming using hevc,'' in \emph{Image Processing
  (ICIP), 2015 IEEE International Conference on}.\hskip 1em plus 0.5em minus
  0.4em\relax IEEE, 2015, pp. 2244--2248.

\bibitem{niamut2016mpeg}
O.~A. Niamut, E.~Thomas, L.~D'Acunto, C.~Concolato, F.~Denoual, and S.~Y. Lim,
  ``Mpeg dash srd: spatial relationship description,'' in \emph{Proceedings of
  the 7th International Conference on Multimedia Systems}.\hskip 1em plus 0.5em
  minus 0.4em\relax ACM, 2016, p.~5.

\bibitem{le2016tiled}
J.~Le~Feuvre and C.~Concolato, ``Tiled-based adaptive streaming using
  mpeg-dash,'' in \emph{Proceedings of the 7th International Conference on
  Multimedia Systems}.\hskip 1em plus 0.5em minus 0.4em\relax ACM, 2016, p.~41.

\bibitem{inoue2010interactive}
M.~Inoue, H.~Kimata, K.~Fukazawa, and N.~Matsuura, ``Interactive panoramic
  video streaming system over restricted bandwidth network,'' in
  \emph{Proceedings of the 18th ACM international conference on
  Multimedia}.\hskip 1em plus 0.5em minus 0.4em\relax ACM, 2010, pp.
  1191--1194.

\bibitem{rondao2012interactive}
P.~Rondao~Alface, J.-F. Macq, and N.~Verzijp, ``Interactive omnidirectional
  video delivery: A bandwidth-effective approach,'' \emph{Bell Labs Technical
  Journal}, vol.~16, no.~4, pp. 135--147, 2012.

\bibitem{yu2015framework}
M.~Yu, H.~Lakshman, and B.~Girod, ``A framework to evaluate omnidirectional
  video coding schemes,'' in \emph{2015 IEEE International Symposium on Mixed
  and Augmented Reality (ISMAR)}.\hskip 1em plus 0.5em minus 0.4em\relax IEEE,
  2015, pp. 31--36.

\bibitem{zakharchenko2016quality}
V.~Zakharchenko, K.~P. Choi, and J.~H. Park, ``Quality metric for spherical
  panoramic video,'' in \emph{SPIE Optical Engineering+ Applications}.\hskip
  1em plus 0.5em minus 0.4em\relax International Society for Optics and
  Photonics, 2016, pp. 99\,700C--99\,700C.

\end{thebibliography}
}

\end{document}